\documentclass[twoside,prb,letterpaper,preprintnumbers,superscriptaddress,notitlepage,onecolumn]{revtex4}

\usepackage{mathptmx}
\usepackage{helvet}
\usepackage{graphicx,fancyhdr}
\usepackage{dashbox,color}%
\usepackage{bm}
\usepackage{sidecap}%
\definecolor{lightblue}{rgb}{0.6,0.9,1}
\definecolor{myrefblue}{rgb}{0.1,0.6,1}
\definecolor{myblue}{rgb}{0,0,0}
\definecolor{nmat}{rgb}{0.7,0.04,0.26}

\newcommand{\nfcaopt}{$\textrm{NaFe}_{0.978}\textrm{Co}_{0.022}\textrm{As}$}

\newcommand{\nfca}{$\textrm{NaFe}_{1-x}\textrm{Co}_{x}\textrm{As}$}

\begin{document}
\pagestyle{fancy}

\renewcommand{\headrule}{\vskip-3pt\hrule width\headwidth height\headrulewidth \vskip-\headrulewidth}

\fancypagestyle{plainfancy}{%
\lhead{}
\rhead{}
\chead{}
\lfoot{}
\cfoot{}
\rfoot{\bf\scriptsize\textsf{\thepage}}
\renewcommand{\headrulewidth}{0pt}
\renewcommand{\footrulewidth}{0pt}
}

\fancyhead[LE,RO]{}
\fancyhead[LO,RE]{}
\fancyhead[C]{}
\fancyfoot[LO,RE]{}
\fancyfoot[C]{}
\fancyfoot[LE,RO]{\bf\scriptsize\textsf{\thepage}}

\renewcommand\bibsection{\section*{\sffamily\bfseries\normalsize {References}\vspace{-10pt}\hfill~}}
\newcommand{\mysection}[1]{\section*{\sffamily\bfseries\normalsize {#1}\vspace{-10pt}\hfill~}}
\renewcommand{\subsection}[1]{\noindent{\bfseries\normalsize #1}}
\renewcommand{\bibfont}{\fontfamily{ptm}\footnotesize\selectfont}
\renewcommand{\figurename}{Figure}
\renewcommand{\refname}{References}
\renewcommand{\bibnumfmt}[1]{#1.}

\makeatletter
\long\def\@makecaption#1#2{%
  \par
  \vskip\abovecaptionskip
  \begingroup
   \small\rmfamily
   \sbox\@tempboxa{%
    \let\\\heading@cr
    \textbf{#1\hskip1pt$|$\hskip1pt} #2%
   }%
   \@ifdim{\wd\@tempboxa >\hsize}{%
    \begingroup
     \samepage
     \flushing
     \let\footnote\@footnotemark@gobble
     \textbf{#1\hskip1pt$|$\hskip1pt} #2\par
    \endgroup
   }{%
     \global \@minipagefalse
     \hb@xt@\hsize{\hfil\unhbox\@tempboxa\hfil}%
   }%
  \endgroup
  \vskip\belowcaptionskip
}%
\makeatother

\thispagestyle{plainfancy}

\fontfamily{helvet}\fontseries{bf}\selectfont
\mathversion{bold}
\begin{widetext}
\begin{figure}
\vskip0pt\noindent\hskip-0pt
\hrule width\headwidth height\headrulewidth \vskip-\headrulewidth
\hbox{}\vspace{4pt}
\hbox{}\noindent\vskip10pt\hbox{\noindent\huge\sffamily\textbf{Weak-coupling superconductivity in a strongly}}\vskip0.05in\hbox{\noindent\huge\sffamily\textbf{correlated iron pnictide}}
\vskip10pt
\hbox{}\noindent\begin{minipage}{\textwidth}\flushleft
\renewcommand{\baselinestretch}{1.2}
\noindent\hskip-10pt\large\sffamily A.~Charnukha$^{1,2,3}$, K.~W.~Post$^1$, S.~Thirupathaiah$^{2,4}$, D.~Pr\"opper$^3$, S.~Wurmehl$^2$, M.~Roslova$^{2,5}$, I.~Morozov$^{2,5}$, B.~B\"uchner$^2$, A.~N.~Yaresko$^3$, A.~V.~Boris$^3$, S.~V.~Borisenko$^2$  \&~D.~N.~Basov$^1$
\end{minipage}
\end{figure}
\end{widetext}

\begin{figure}[!h]
\begin{flushleft}
{\footnotesize\sffamily
$^1$Physics Department, University of California--San~Diego, La Jolla, CA 92093, USA, $^2$Leibniz Institute for Solid State and Materials Research, IFW, 01069 Dresden, Germany, $^3$Max Planck Institute for Solid State Research, 70569 Stuttgart, Germany, $^4$Solid State and Structural Chemistry Unit, Indian Institute of Science, Bangalore--560 012, India, $^5$Department of Chemistry, Moscow State University, 119991 Moscow, Russia. $^\dagger$ E-mail: acharnukha@ucsd.edu.}
\end{flushleft}
\end{figure}

\fontsize{8pt}{8pt}\selectfont
\renewcommand{\baselinestretch}{0.9}
\noindent\sffamily\bfseries{Iron-based superconductors have been found to exhibit an intimate interplay of orbital, spin, and lattice degrees of freedom, dramatically affecting their low-energy electronic properties, including superconductivity. Albeit the precise pairing mechanism remains unidentified, several candidate interactions have been suggested to mediate the superconducting pairing, both in the orbital and in the spin channel. Here, we employ optical spectroscopy (OS), angle-resolved photoemission spectroscopy (ARPES), {\it ab initio} band-structure, and Eliashberg calculations to show that nearly optimally doped \nfcaopt\ exhibits some of the strongest orbitally selective electronic correlations in the family of iron pnictides. Unexpectedly, we find that the mass enhancement of itinerant charge carriers in the strongly correlated band is dramatically reduced near the $\Gamma$~point and attribute this effect to orbital mixing induced by pronounced spin-orbit coupling. Embracing the true band structure allows us to describe all low-energy electronic properties obtained in our experiments with remarkable consistency and demonstrate that superconductivity in this material is rather weak and mediated by spin fluctuations.}
\mathversion{normal}
\normalfont\normalsize

Strong entanglement of various electronic and lattice degrees of freedom in the iron-based superconductors has become the {\it leitmotif} of recent condensed-matter research and has been identified with a vast variety of experimental probes~\cite{Yi26042011,Fisher_BFCA_Nematicity_2012,PhysRevLett.111.217001,ParkDaiNematicityINS_2014} and interpreted theoretically~\cite{NematicityPnictides_Fernandes_NatPhys2014,PhysRevLett.111.047004}. This inherent complexity arises from multiple partially filled Fe-$3d$ orbitals simultaneously contributing to the low-energy quasiparticle dynamics and strongly affected by the Hund's coupling correlations~\cite{PhysRevLett.112.177001}. The orbitally selective nature of these interactions leads to a likewise orbitally selective bandwidth renormalization and relative energy shift of various bands with respect to each other and the Fermi level due to the pronounced particle-hole asymmetry of the electronic structure~\cite{Cappelluti2010S508}. Such non-trivial interaction-induced modifications result in a singular Fermi-surface topology that is dramatically different from the predictions of theoretical calculations in both the number and the binding energy of the bands crossing the Fermi level~\cite{Charnukha_SFCAO_2015,Borisenko_BKFA_FS_NormState2009}, strongly affecting low-energy electronic properties including superconductivity.

Here, we show that in the nearly optimally doped \nfcaopt\ compound, low-energy orbitally selective renormalization of the band structure deviates markedly from the general experimental and theoretical renormalization trend in iron-based superconductors~\cite{YinHauleKotliar_HundsCouplingAndRenorm_2011}. We find more than twice stronger than expected orbitally selective electronic correlations in the Fe-$3d_\mathrm{xy}$ band crossing the Fermi level near the $\Gamma$ point of the Brillouin zone, similarly to $\textrm{FeTe}_{1-x}\textrm{Se}_x$~\cite{PhysRevLett.104.097002,PhysRevLett.110.037003,PhysRevB.89.220506}. However, in contrast to the observations in the latter compounds, the mass enhancement in the strongly correlated band of \nfcaopt\ is drastically reduced near the Fermi level, where our relativistic {\it ab initio} calculations predict orbital mixing of the Fe-$3d_{xy}$,~$3d_{xz}$, and~$3d_{yz}$ orbitals to occur due to spin-orbit coupling, leading to a pronounced modification of low-energy electronic properties. A simultaneous detailed analysis of the ARPES and OS data reveals remarkable consistency in the extracted itinerant properties of both the normal and the superconducting state. It allows us to conclude that the ubiquitous dichotomy between the coherent and incoherent far-infrared quasiparticle response observed in all iron-based superconductors~\cite{0953-8984-26-25-253203} is not dominated by the disparity in the mobilities of the hole and electron charge carriers but rather originates in the inelastic scattering from an intermediate bosonic excitation. Finally, all of the observed features in the far-infrared conductivity in the superconducting and normal state can be reproduced in an effective two-band Eliashberg model assuming relatively weak superconducting pairing with $s_\pm$ symmetry mediated by low-energy spin fluctuations with the frequency and temperature dependence observed in the same compound by inelastic neutron scattering~\cite{PhysRevB.88.064504}.

\begin{figure}[!b]
\vskip50pt
\end{figure}

\begin{figure*}[!t]
\includegraphics[width=\textwidth]{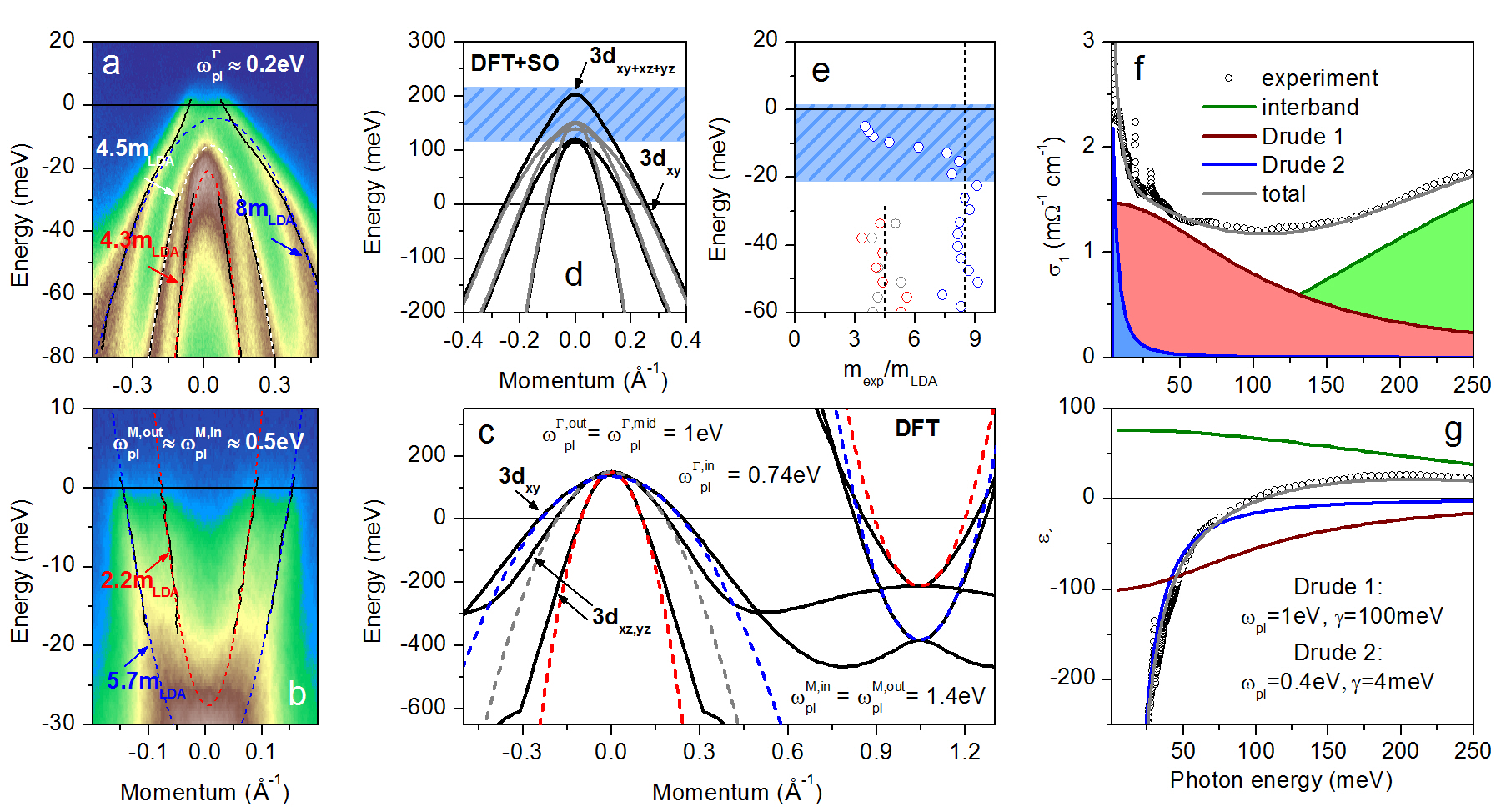}
\caption{\label{fig:renorm}\textbf{Low-energy electronic properties of \nfcaopt.}~\textbf{a,b,} Low-energy electronic structure of \nfcaopt\ in the normal state near the $\Gamma$ (\textbf{a}, $30\ \textrm{K}$) and M~(\textbf{b}, $22\ \textrm{K}$) points of the Brillouin zone recorded using $20\ \textrm{eV}$ and $25\ \textrm{eV}$ photons linearly polarized perpendicular to and within the plane of incidence, respectively. The dispersion of all electronic bands observed in experiment (black solid lines) has been obtained using a multi-Lorentzian fit of the momentum-distribution curves. Effective-mass renormalization indicated in the panels has been calculated by comparing the low-energy parabolic fits to the electronic dispersions extracted from the experimental data (dashed lines in panels~\textbf{a,b}) with those to the corresponding electronic bands predicted theoretically (dashed lines in panel \textbf{c}).~\textbf{c,} Theoretically predicted low-energy electronic band structure of NaFeAs in the $\Gamma-M$ high-symmetry direction of the Brillouin zone (black solid lines) and low-energy parabolic fits to the dispersions near $\Gamma$ and~M (dashed lines). The colors of the dashed lines correspond to those in panels~\textbf{a,b}. Plasma frequencies of all bands crossing the Fermi level extracted from experimental data and predicted by theory are shown in panels~\textbf{a--c}.~\textbf{d,} Electronic band structure in the $\Gamma-M$ high-symmetry direction of the Brillouin zone near the $\Gamma$ point obtained in a relativistic DFT calculation (black lines), which explicitly accounts for a finite spin-orbit coupling (its characteristic energy scale is shown as blue hatched area). The results of the corresponding non-relativistic calculation from panel~\textbf{c} are shown for comparison (grey lines).~\textbf{e,} Energy dependence of the effective mass renormalization in the outer (blue open circles), middle (grey circles), and inner (red circles) hole band at the $\Gamma$~point with respect to the corresponding low-energy quasiparticle masses extracted from the theoretically predicted band structure in panel~\textbf{c} using parabolic fits. The energy scale affected by spin-orbit interaction is indicated as a blue hatched area.~\textbf{f,g,} Real part of the optical conductivity (\textbf{f}) and the dielectric function (\textbf{g}) of \nfcaopt\ obtained in OS measurements (open circles) along with the results of a Drude-Lorentz dispersion analysis (blue and red lines/shaded areas and green line/shaded area for the two Drude terms and the total contribution of all interband transitions, respectively; gray solid lines indicate the sum of all terms). The parameters of the itinerant charge carrier response obtained by means of this analysis are summarized in panel~\textbf{g}.}
\end{figure*}

The effect of electronic correlations on the low-energy electronic structure of \nfcaopt\ is illustrated in Figs.~\ref{fig:renorm}a--c. The general band structure of \nfca\ has been studied extensively and consists, similarly to most iron-based superconductors, of three hole bands at the $\Gamma$~(Z) point of the Brillouin zone and two electron bands near the M~point of the two-Fe Brillouin zone~\cite{PhysRevB.86.214508,PhysRevB.84.064519} (the difference in the band structure at the $\Gamma$ and Z~points is minor, see Supplementary Information). Both of these sets of electronic dispersions are analyzed in Figs.~\ref{fig:renorm}a,b, respectively, and compared to the corresponding predictions of our band-structure calculations shown in Fig.~\ref{fig:renorm}c. In stark contrast to the prediction of the theory in Fig.~\ref{fig:renorm}c, only one hole band at the $\Gamma$ point of \nfcaopt\ crosses the Fermi level, the remaining two are shifted to lower binding energies and terminate within about $20\ \textrm{meV}$ below the Fermi level. These inner hole bands have a well-defined parabolic shape, as illustrated by the fits (lower white and red dashed lines) to the experimental band dispersions extracted from momentum-distribution curves at various binding energies (black solid lines). The comparison of the fitted effective masses to the theoretical band structure in Fig.~\ref{fig:renorm}c reveals a renormalization by a factor of 4.3 and 4.5 for the inner and middle hole bands of Fe-$3d_{xz,yz}$ orbital character, respectively. The outer hole band of Fe-$3d_{xy}$ orbital character in Fig.~\ref{fig:renorm}a shows a dramatic renormalization by a factor of more than $8$ (blue dashed line), clearly demonstrating the existence of very strong and orbitally selective electronic correlations in \nfcaopt, similar to $\textrm{FeTe}_{1-x}\textrm{Se}_x$~\cite{PhysRevLett.104.097002,PhysRevLett.110.037003,PhysRevB.89.220506} (the slight asymmetry of the photoemission intensity distribution in Fig.~\ref{fig:renorm}a does not affect our conclusions, see Supplementary Information). Such a strong enhancement of the effective quasiparticle mass is at odds with the predicted renormalization by a factor of only $3$ to $4$, identified in combined density-functional and dynamic mean-field theory calculations (DFT+DMFT) in Refs.~\onlinecite{YinHauleKotliar_HundsCouplingAndRenorm_2011}. Our analogous analysis of the experimental band structure of \nfcaopt\ at the M~point of the Brillouin zone (Fig.~\ref{fig:renorm}d) reveals two well-defined parabolic bands of electron character, consistent with the prediction of theory in Fig.~\ref{fig:renorm}c, with a very disparate effective-mass renormalization of about 2.2 and 5.7 for the inner and outer electron bands, respectively.

Quite surprisingly, the outer hole band's dispersion reveals a departure from a parabolic shape (blue dashed line in Fig.~\ref{fig:renorm}a) upon approaching the Fermi level, clearly absent in the calculated band structure. To quantify this non-parabolicity, we extract the energy-dependent effective mass $m(E)=\hbar^2k(E)dk/dE$, where $\hbar$ is the Planck constant, $k$ --- quasiparticle wave vector, and $E$ --- quasiparticle energy. For a parabolic dispersion $m(E)=const$. The resulting energy-dependent effective mass renormalization $m_\mathrm{exp}(E)/m_\mathrm{LDA}$ in all hole bands near the $\Gamma$ point of the Brillouin zone ($m_\mathrm{LDA}$ from Fig.~\ref{fig:renorm}c) is shown in Fig.~\ref{fig:renorm}e. It is clear that while the higher-energy (outside of the blue hatched area) effective mass of the outer hole band is simply renormalized from its theoretical counter part by a factor of~8, at low energies this is not the case.

This dramatic reduction of the effective mass at lower binding energies cannot be attributed solely to the effect of high-energy electronic correlations, which typically lead to an enhancement of the effective mass over the entire electronic band width. In the presence of orbitally selective renormalization the same is true for each band of a given orbital character. Our {\it ab initio} calculations (Fig.~\ref{fig:renorm}c) clearly show that the outer hole band has a well-defined $3d_{xy}$ orbital character up to its termination. Recently, it has been demonstrated that spin-orbit coupling is not negligible in iron-based superconductors~\cite{2014arXiv1409.8669B}. Provided that the hole bands of $3d_{xy}$ and $3d_{xz,yz}$ character are almost degenerate at the $\Gamma$ point of the Brillouin zone, as can be seen in Fig.~\ref{fig:renorm}c, the existence of a sizable spin-orbit interaction would lead to the mixing of the orbital character in these three bands and potentially reduce the mass renormalization in the outer $3d_{xy}$ hole band due to a large admixture of more weakly renormalized $3d_{xz,yz}$ bands. The existence of such orbital mixing has been identified previously based on the polarization dependence of the photoemission intensity near the $\Gamma$~(Z) point in Ref.~\onlinecite{PhysRevB.86.214508}. Given the reduction of the effective mass in the outer hole band due to orbital mixing, one should expect a corresponding enhancement of the effective mass in the middle and/or inner hole bands due to the finite admixture of the heavier $3d_{xy}$ orbital. This effect, albeit rather subtle, can nevertheless be clearly identified in the second derivative of the experimental data in Fig.~\ref{fig:renorm}a upon close inspection (see Supplementary Information).

The aforementioned reduction of the effective mass due to orbital mixing does not rely on the existence of strong electronic correlations and should be observable already at the level of quasiparticle band masses. Indeed, Fig.~\ref{fig:renorm}d demonstrates that when spin-orbit coupling in this material is taken into account in a relativistic {\it ab initio} calculation (black solid lines), the dispersion of the hole bands at low energies shows a pronounced departure from that obtained in the non-relativistic calculation presented in Fig.~\ref{fig:renorm}c (grey solid lines). Both the reduction of the effective mass in the outer band and its enhancement in the middle and inner bands are apparent.

\begin{figure*}[!t]
\includegraphics[width=\textwidth]{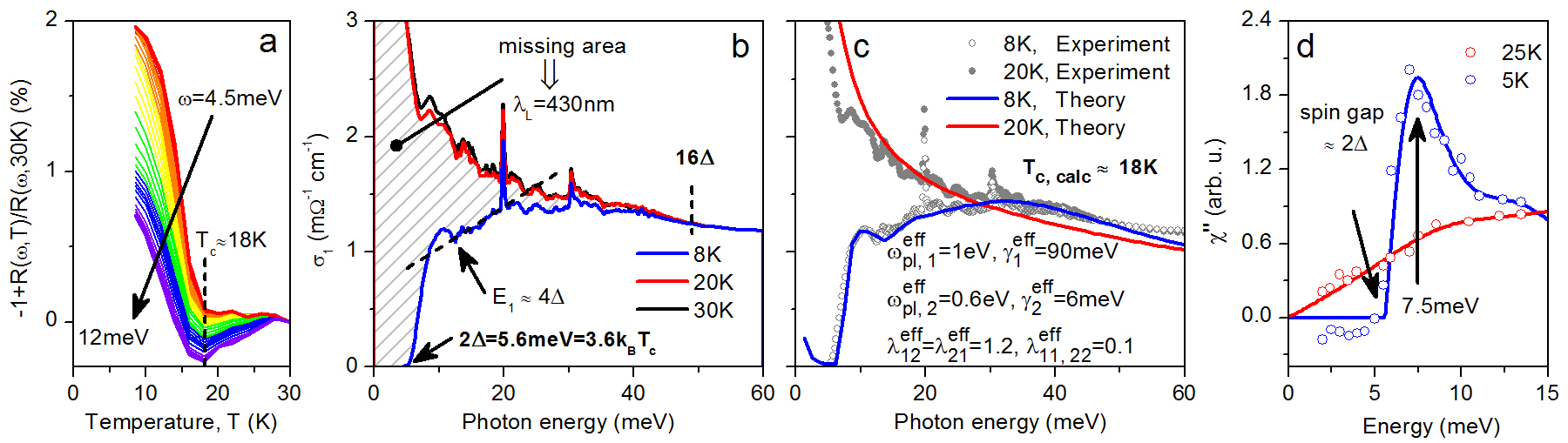}
\caption{\label{fig:sc}\textbf{Superconductivity in nearly optimally doped \nfcaopt.}~\textbf{a,} Relative change in the sample reflectance as a function of temperature for several frequencies in the THz spectral range. The transition into the superconducting state is evidenced by a marked increase in the sample reflectance below $T_{\mathrm{c}}\approx18\ \textrm{K}$ (vertical dashed line).~\textbf{b,} Real part of the itinerant THz and far-infrared conductivity of \nfcaopt\ in the superconducting state at $8\ \textrm{K}$ (blue solid line) and normal state at $20$ and $30\ \textrm{K}$ (red and black solid lines, respectively). The formation of Cooper pairs is manifested in the missing area under the conductivity curve in the superconducting vs normal state (hatched area), which amounts to a London penetration depth of $430\ \textrm{nm}$. Optical superconducting energy gap $2\Delta$ is clearly visible as the energy at which the real part of the optical conductivity vanishes. The frequency-dependence of the optical conductivity above the initial onset of absorption in the superconducting state exhibits a quasilinear character, which defines an additional energy scale, $E_1$ (for interpretation see text). Normal state optical conductivity is recovered at about $16\Delta$ (vertical dashed line).~\textbf{c,} Comparison of the real part of the optical conductivity in the superconducting ($8\ \textrm{K}$, gray open circles) and normal ($20\ \textrm{K}$, gray filled circles) state obtained experimentally to the results of theoretical calculations in the framework of an effective two-band Eliashberg theory at $8\ \textrm{K}$ (blue solid line) and $20\ \textrm{K}$ (red solid line). Absorption below $2\Delta$ in the superconducting state is due to thermally excited quasiparticles. All parameters of the theory are indicated in the panel. The initial sharp onset of absorption at $2\Delta$ is assisted by elastic impurity scattering while the quasilinear frequency dependence of the optical conductivity at higher energies results from Holstein processes (breaking of a Cooper pair with a simultaneous creation of one or several quanta of the mediating boson).~\textbf{d,} The spectral function of the mediating boson in the normal ($25\ \textrm{K}$) and superconducting ($5\ \textrm{K}$) state used in the calculation (red and blue solid lines, respectively) and the imaginary part of the spin susceptibility obtained using inelastic neutron scattering on the same compound (open circles; data from Ref.~\onlinecite{PhysRevB.88.064504}). Imaginary part of the spin susceptibility in the superconducting state clearly exhibits a resonance mode at $7.5\ \textrm{meV}$ and a spin gap (SG) of $5.6\ \textrm{meV}$, approximately equal to the optical superconducting energy gap $2\Delta$ in this material.}
\end{figure*}

The complex modification of the low-energy electronic structure with respect to theory observed in our experiment has a profound effect on the itinerant properties of \nfcaopt. It strongly affects the analysis and interpretation of the OS data obtained on the same compound, as we will demonstrate below. The itinerant quasiparticle response can be extracted from the OS data by eliminating the contribution of all clearly identifiable interband transitions (green lines in Figs.~\ref{fig:renorm}f,g) from the optical conductivity $\sigma(\omega)$ or, equivalently, dielectric function $\varepsilon(\omega)=1+4\pi i\sigma(\omega)/\omega$ (see also Supplementary Information). The resulting itinerant optical response $\sigma^\mathrm{it}(\omega)$, or $\varepsilon^\mathrm{it}(\omega)$, can be decomposed into a narrow (coherent) and a broad (incoherent) Drude-like contribution (blue and red lines in Figs.~\ref{fig:renorm}f,g). Such a decomposition is rather common in the iron-based superconductors and has been observed in essentially all known materials of this family~\cite{0953-8984-26-25-253203}. At the same time, the electronic band structure of \nfcaopt\ features three bands crossing the Fermi level, indicating the existence of a segregation of all free charge carriers into two {\it effective} electronic subsystems: one with a large and the other one with a small scattering rate (summarized in Fig.~\ref{fig:renorm}g). 

It is tempting to assign the narrow Drude component to the combined response of charge carriers on the electron sheets of the Fermi surface (which in iron-based compounds have been found to have higher mobilities than hole carriers based on Hall and quantum-oscillation measurements~\cite{PhysRevB.80.140508,PhysRevLett.101.216402}), and the broad one to the contribution of the carriers on the hole sheet of the Fermi surface. However, a direct comparison of the plasma frequencies of the two Drude terms extracted in our analysis of the optical conductivity (Fig.~\ref{fig:renorm}g) to those obtained from the ARPES data (as shown in Figs.~\ref{fig:renorm}a,b and detailed in the supplementary information) reveals that this assignment is incorrect: the plasma frequency of the hole sheet of the Fermi surface, $\omega_\mathrm{pl}^{\mathrm{hole}}=0.2\ \textrm{eV}$, is substantially smaller than the total plasma frequency of the electron sheets, $\omega_\mathrm{pl,tot}^{\mathrm{el}}=\sqrt{\omega_\mathrm{pl}^{\mathrm{el,in}}+\omega_\mathrm{pl}^{\mathrm{el,out}}}=0.7\ \textrm{eV}$. The reverse assignment (broad Drude component to electron charge carriers) cannot be reconciled with the negative Hall coefficient observed in this compound~\cite{1367-2630-15-4-043048}. Additionally, a simple estimate of the electron mean free path due to elastic impurity scattering as $l_0=v_\mathrm{F}/\gamma$, where $v_\mathrm{F}$ is the Fermi velocity and $\gamma$ is the quasiparticle scattering rate obtained in our Drude-Lorentz analysis, leads to an unrealistic result. Taking $v_\mathrm{F}=0.6\ \textrm{eV}\text{\AA}$ for the electron bands in Fig.~\ref{fig:renorm}b and $\gamma=0.1\ \textrm{eV}$ for the broad Drude component in Figs.~\ref{fig:renorm}f,g, one arrives at $l_0=6$~\AA\ or less than two unit cells. Such a small mean free path is hard to justify in a nearly stoichiometric \nfcaopt. A similar estimate assuming $\gamma=0.1\ \textrm{eV}$ on the hole pocket ($v_\mathrm{F}=0.15\ \textrm{eV}\text{\AA}$) leads to $l_0=1.5$~\AA, or less than one unit cell. Analogous analysis of the broad Drude component in the optical conductivity of optimally doped $\textrm{BaFe}_{1.85}\textrm{Co}_{0.15}\textrm{As}_2$ previously extracted a comparable mean free path assuming elastic scattering of holes~\cite{PhysRevB.82.174509}. These inconsistencies imply that the large scattering rate of the incoherent Drude component does not originate in the low mobility of one of the charge-carrier type but rather results from {\it inelastic} scattering of charge carriers (of either sign) from an intermediate boson. This conclusion is consistent with the association of the broad Drude component with electronic correlations based on the doping dependence of the optical conductivity in various iron-based compounds~\cite{PhysRevB.88.094501,NakajimaUchidaOpticsBFA2014}.

The strength of electronic correlations is frequently estimated by comparing the total itinerant spectral weight (area under the itinerant optical conductivity curve) $SW=\int_0^\infty\sigma_1^\mathrm{it}(\Omega)d\Omega=\left(\omega_{\mathrm{pl,tot}}^{\mathrm{it, exp}}\right)^2/8$, where $\omega_{\mathrm{pl,tot}}^{\mathrm{it,exp}}$ is the total itinerant plasma frequency, to the predictions of theoretical band-structure calculations~\cite{RevModPhys.77.721}. While in particle-hole--symmetric single-band materials this approach provides a good estimate of effective-mass renormalization and thereby correlations, in iron-based superconductors the experimentally observed plasma frequency is modified from its theoretical prediction not only by orbitally selective band width renormalization but also intrinsic (see Supplementary Information) band shifts~\cite{Charnukha_SFCAO_2015}, a process that in general conserves neither the Fermi wave vector nor the number of bands crossing the Fermi level, dramatically affecting the total plasma frequency. Therefore, the assessment of electronic correlations in such cases can only be reliably carried out based on the comparison of the complete low-energy experimentally observed {\it band structure} to its theoretically predicted counterpart.

Fortunately, even in such complicated cases the spectral-weight analysis of the itinerant optical response bears fruit. It is well-known that the total spectral weight of any system (including optical absorption in the ultraviolet and x-ray spectral range) is constant and does not depend on the details of the band structure: $SW^\mathrm{tot}=\int_0^{\infty}\sigma^\mathrm{tot}(\Omega)d\Omega=4\pi ne^2/m_e$ (in CGS units), where $n$ is the electron density, $e$ and $m_e$ are the bare electron charge and mass, respectively. Quite similarly, the {\it itinerant} spectral weight (devoid of all interband transitions) is also constant, with $m_e$ substituted by the quasiparticle band mass including renormalization at energies larger than the plasma frequency. Any low-energy mass renormalization due to, e.g., coupling to phonons or other bosonic excitations will not affect this spectral weight as long as its characteristic energy is smaller than the plasma frequency of free charge carriers. On the other hand, the low-energy band structure and the plasma frequency extracted from it will be affected by such renormalization. Therefore, the comparison of the total itinerant plasma frequency obtained from OS and ARPES data provides access to purely boson-exchange--induced effective mass renormalization even when it is not immediately apparent in the experimental band structure. In the present case of \nfcaopt\ this renormalization amounts to $m_\mathrm{OS}/m_\mathrm{ARPES}=(\omega_{\mathrm{pl,tot}}^{\mathrm{it,OS}}/\omega_{\mathrm{pl,tot}}^{\mathrm{ARPES}})^2=2.15$ (see Fig.~\ref{fig:renorm}a,b,g). This is related to the strength of electron-boson couling $\lambda$ via $1+\lambda=m_\mathrm{OS}/m_\mathrm{ARPES}$, giving $\lambda=1.15$.

Having established that only three out of five electronic bands predicted theoretically actually contribute to the itinerant carrier response of \nfcaopt\ and that electrons are coupled to an intermediate boson with a coupling strength of $\lambda=1.15$, we now turn to the analysis of the superconducting state that develops in this electronic structure. Up to now, the character of superconductivity in optimally doped \nfca\ has remained unclear, with both weak and strong superconducting pairing suggested based on the ARPES measurements of the superconducting energy gap in Refs.~\onlinecite{PhysRevB.86.214508,PhysRevB.84.064519}, respectively. This uncertainty can be eliminated based on the analysis of our bulk OS measurements in the far-infrared and THz spectral range, shown in Fig.~\ref{fig:sc}. Panel~\ref{fig:sc}a shows the temperature dependence of the relative change of the reflectance with respect to that at $30\ \textrm{K}$ for several frequencies in the lowest THz spectral range. The normal-state trend for all frequencies is suddenly reversed near $18\ \textrm{K}$, which thus must be identified with the superconducting transition temperature, consistent with the previous ARPES measurements on the same samples~\cite{PhysRevB.86.214508}. The analysis of the real part of the optical conductivity in the far-infrared and THz spectral range reveals the quintessential feature of Cooper pairing --- the dramatic suppression of the optical conductivity, with the missing area $\delta A_{\mathrm{L}}$ between the normal and the superconducting state at finite frequencies $\omega>0$ ($\delta A_{\mathrm{L}} =\int_{0^+}^{16\Delta}(\sigma_1^{\mathrm{20K}}(\omega)-\sigma_1^{\mathrm{8K}}(\omega))d\omega$, hatched area in Fig.~\ref{fig:sc}b) transferred into the coherent response of the condensate at $\omega=0$ (Ref.~\onlinecite{Tinkham_superconductivity_1995_articlestyle}). Our analysis shows that this missing area corresponds to a London penetration depth $\lambda_\mathrm{L}=c/\sqrt{8\delta A_\mathrm{L}}=430\ \textrm{nm}$, in a remarkable agreement with nuclear-magnetic--resonance and muon-spin--rotation measurements on this material~\cite{PhysRevB.87.174517,PhysRevLett.104.057007}.

The vanishing of the optical conductivity in the superconducting state (blue solid line in Fig.~\ref{fig:sc}b) further identifies the optical superconducting energy gap $2\Delta\approx5.6\ \textrm{meV}$ and the corresponding gap ratio $2\Delta/k_{\mathrm{B}}T_{\mathrm{c}}\approx3.6$, very close to the weak-coupling limit of the Bardeen-Cooper-Schrieffer theory of superconductivity~\cite{PhysRev.106.162,Tinkham_superconductivity_1995_articlestyle}. These values are in a good agreement with those extracted from the ARPES data on the same single crystals~\cite{PhysRevB.86.214508} and together provide compelling evidence for relatively weak-coupling superconductivity in \nfcaopt.

The onset of absorption at $2\Delta$ (blue solid line in Fig.~\ref{fig:sc}b) is very sharp and has a Mattis-Bardeen shape, characteristic of a weakly coupled superconductor or that with large elastic impurity scattering~\cite{PhysRev.111.412,PhysRevLett.3.552}. Intriguingly, the frequency dependence of the optical conductivity beyond this initial increase is non-monotonic and reveals a quasilinear section, which deviates markedly from the Mattis-Bardeen shape and defines another energy scale, $E_1$. Such a quasilinear frequency behavior in the optical conductivity of iron-based superconductors is not unprecedented and has been previously observed in the response of optimally doped $\textrm{(Ba,K)Fe}_2\textrm{As}_2$~\cite{PhysRevLett.101.107004,CharnukhaNatCommun2011} and attributed to the boson-mediated absorption in the clean limit (low impurity scattering) of a strongly coupled superconductor~\cite{PhysRevB.84.174511}.

It is quite clear that the combination of such very different features characteristic of a weakly and strongly coupled as well as clean- and dirty-limit superconductor in the optical response of the same material necessitates a comparison with a theoretical model that encompasses all of these limiting behaviors. The Eliashberg theory of superconductivity is perfectly suited to this task. The key input parameter of the theory is the spectrum of the bosonic excitation that mediates superconducting pairing and the overall electron-boson coupling strength. Spin dynamics in \nfcaopt\ has been shown to be sensitive to superconductivity, as manifested in the formation of a resonance peak in the imaginary part of the spin susceptibility below $T_\mathrm{c}$ (Refs.~\onlinecite{WenSTM_BKFA_2012_NatPhys,PhysRevB.88.064504}), while the electron-phonon interaction has been demonstrated to be insufficient to account for the elevated superconducting transition temperature of iron-based superconductors~\cite{boeri:026403}. Therefore, we assume interband electron-electron interaction via the exchange of spin fluctuations with the experimentally obtained spectral shape in the normal and superconducting state (Fig.~\ref{fig:sc}d) to mediate superconducting pairing in our model. 

In the presence of the disparate features observed in the optical response of \nfcaopt\ and an interband pairing interaction, the simplest, minimal, model would incorporate two effective electronic bands. The two-band Eliashberg formalism employed in our calculations has been detailed in Ref.~\onlinecite{1367-2630-15-1-013002}. We assume the symmetry of the superconducting order parameter to be of the $s_\pm$-type. Given that the densities of states of all bands crossing the Fermi level are comparable and that all ARPES measurements find little variation in the size of the superconducting energy gap within and between the bands~\cite{PhysRevB.86.214508,PhysRevB.84.064519}, we assume two effective bands with isotropic superconducting energy gaps of the same magnitude and opposite sign, as well as equal densities of states. Elastic (impurity) {\it interband} scattering is neglected~\cite{PhysRevB.84.174511}. The plasma frequencies, {\it intraband} impurity scattering rates, and the coupling strength between the bands are the free parameters of the model tuned to fit the optical superconducting energy gap $2\Delta=5.6\ \textrm{meV}$, the superconducting transition temperature $T_{\mathrm{c}}=18\ \textrm{K}$, and the overall shape of the optical conductivity in the superconducting and normal state. Their optimal values obtained in the fit under the aforementioned constraints are indicated in Fig.~\ref{fig:sc}c.

Figure~\ref{fig:sc}c shows the result of our modeling (solid lines) overlaid on the experimental data in the superconducting (open circles) and normal (filled circles) state. It is immediately evident that {\it quantitative} agreement can be achieved for the parameters given in the panel. Quite remarkably, if not unprecedentedly, all of the transport parameters (the individual plasma frequencies of the bands and their {\it effective} impurity scattering rates) are in a very good agreement with the properties of the two effective itinerant electronic subsystems extracted in the Drude analysis of the conductivity data in Fig.~\ref{fig:renorm}g. Furthermore, the interband electron-boson coupling strength obtained from the fit ($\lambda_{12}=\lambda_{21}=1.2$) shows excellent agreement with the value obtained from the spectral-weight analysis above ($\lambda=1.15$). 

Our theoretical analysis of the optical conductivity in the framework of an effective two-band Eliashberg model thus reconciles all of the characteristic energy scales and spectral features observed in \nfcaopt. The sharp onset of absorption at $2\Delta$ is dominated by the band with large impurity scattering. Energy scale $E_1$ can be traced back to the onset of absorption in a clean superconductor via a Holstein-type process of breaking-up of a Cooper pair with a simultaneous creation of one spin-fluctuation quantum~\cite{PhysRevB.43.473,PhysRevLett.109.027006}. In our model the lowest energy cost of this process is $2\Delta+SG\approx11.2\ \textrm{meV}$, where $SG\approx5.6\ \textrm{meV}$ is the spin gap observed in the spin-fluctuation spectrum (see Fig.~\ref{fig:sc}d), below which no spin fluctuations exist in the superconducting state. The quasilinear behavior above $E_1$ results from the quasilinear shape of the spin-fluctuation spectrum above the spin gap.

The remarkable agreement between different experimental probes and theoretical tools employed in this work demonstrates the importance of employing a realistic electronic band structure to interpret the low-energy electronic properties in such complex multiband systems as iron-based superconductors, if complete and consistent understanding of their properties is to be achieved. Our observation of a dramatic reduction of the effective mass in one of the electronic bands of \nfcaopt, accompanied by relatively weak superconducting pairing, underlines the intimate interplay between different electronic degrees of freedom and the importance of spin-orbit coupling for the ground-state properties of the iron-based superconductors. It further shows that the modification of the orbital character of the bands crossing the Fermi level by means of intrinsic or extrinsic strain may allow for unprecedented control of low-energy electronic properties of these compounds, including superconductivity.

\footnotesize
\mysection{Methods}
\par Angle-resolved photoemission measurements were performed using synchrotron radiation (``$1^3$--ARPES'' end-station at BESSY) within the range of photon
energies $20$-–$90\ \textrm{eV}$ and various polarizations on cleaved surfaces of high-quality single crystals of \nfcaopt. The overall energy and angular resolution were $\sim 5\ \textrm{meV}$ and $0.3^\circ$, respectively, for the low temperature measurements.
\par Optical-spectroscopy measurements were carried out in the THz and the far-infrared spectral range by means of the reflectance technique and from the far-inrared to the ultraviolet spectral range using spectroscopic ellipsometry. The air-sensitive samples of optimally doped \nfcaopt\ were cleaved before every measurement and loaded into the measurement chamber in the neutral atmosphere of argon gas, without exposing the sample to air at any time. High-accuracy absolute measurements of the sample's reflectance were carried out using the gold-overfilling technique.
\par High-quality single crystals of superconducting \nfcaopt\ were synthesized by the self-flux technique and were characterized by x-ray diffraction, transport and magnetization measurements~\cite{PhysRevB.91.184516}. The latter revealed a superconducting transition temperature of about $18\ \textrm{K}$. The width of the superconducting transition was found to be less than $1\ \textrm{K}$, indicating very high homogeneity of the investigated samples.
\par Band structure calculations were performed for the experimental crystal structure of NaFeAs (Ref.~\onlinecite{B818911K}) using the PY~LMTO computer code \cite{Yaresko_LDA_2004_articlestyle}.


\mysection{Acknowledgements}
\noindent We would like to thank O.~V.~Dolgov and A.~A.~Golubov for providing the code used for the calculation of the optical conductivity in the Eliashberg formalism and B.~Keimer for stimulating discussions. A.C. acknowledges financial support by the Alexander von Humboldt foundation. This project was supported by the German Science Foundation under Grant No. BO~1912/2-2, through the Emmy Noether Programme in project WU595/3-1 and WU595/3-2 (S.W.) and within SPP~1458  (BU887/15-1). S.W. thanks the BMBF for support in the framework of the ERA.Net RUS project (project 01DJ12096, Fe-SuCo).  S.T. appreciates financial support by Department of Science and Technology through INSPIRE-Faculty fellowship (Grant No.~IFA-14~PH-86). We would like to thank A.~Wolter-Giraud, F.~Steckel, and C.~Hess for sample characterization.
\mysection{Author contributions}
\noindent A.C., K.W.P., S.T., and D.P. carried out the experiments. A.C. analyzed the data and wrote the manuscript. S.W., M.R., and I.M. carried out the sample growth and characterization. A.N.Y. performed {\it ab initio} calculations. B.B., A.V.B., S.V.B., and D.N.B. supervised the project. All authors discussed the results and reviewed the manuscript.
\mysection{Additional information}
\noindent {\bf Competing financial interests:} The authors declare no competing financial interests.\\
\noindent Correspondence should be addressed to A.C. (acharnukha@ucsd.edu) and D.N.B. (dbasov@physics.ucsd.edu). Requests for materials should be addressed to S.W. (s.wurmehl@ifw-dresden.de).
\eject\newpage

\setcounter{figure}{0}
\renewcommand{\bibnumfmt}[1]{#1.}
\renewcommand{\bibnumfmt}[1]{S#1.}
\renewcommand\thefigure{S\arabic{figure}}
\renewcommand{\cite}[1]{[S\citenum{#1}]}

\thispagestyle{plainfancy}

\fontfamily{helvet}\fontseries{bf}\selectfont
\mathversion{bold}
\begin{widetext}
\begin{figure}
\vskip0pt\noindent\hskip-0pt
\hrule width\headwidth height\headrulewidth \vskip-\headrulewidth
\hbox{}\vspace{4pt}
\hbox{}\noindent\vskip10pt
\begin{center}
\huge\sffamily\textbf{Supplementary information}
\end{center}
\end{figure}
\end{widetext}

\normalfont\normalsize
\section{Spin-orbit coupling in the low-energy electronic structure of \nfcaopt\ near the $\Gamma$ point}
\par In order to confirm that the slight anisotropy of the photoemission intensity in the energy-momentum cut shown in Fig.1a of the main text does not affect the conclusions of our study we have fitted the left and right branches of the outer band's dispersion independently and compared these fits to the results of the simultaneous fit of both branches presented in the text. Only the data points at high enough binding energies (unaffected by the reduction in the effective mass due to the spin-orbit coupling) were used for all fits (red and green filled circles); the entire dispersion of the band is shown for completeness (grey empty circles).

\begin{figure}[b!]
\includegraphics[width=\textwidth]{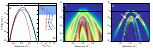}
\caption{\label{fig:zcomp}\textbf{Details of the low-energy electronic band structure of \nfcaopt\ near the $\Gamma$ point.}~\textbf{a,} Energy dispersion of the outer hole band near the $\Gamma$ point of the Brillouin zone (circles) extracted from the maximum of the quasiparticle peak in the momentum-distribution curves overlaid with the simultaneous parabolic fit of both the left and right branches (blue solid line) as well as independent parabolic fits of the left (red solid line) and right (green solid line) branches. The data used for the fitting are shown as filled circles of the corresponding color. Grey open circles show the energy dispersion up to the Fermi level.~\textbf{b,} Renormalization of the effective mass extracted independently from the left (red open circles) and right (green open circles) branches of the dispersion of the outer hole band, as well as their average, as shown in Fig.1e of the main text.~\textbf{c,d,} Energy-momentum cut along the $\Gamma$-M direction of the Brillouin zone plotted in Fig.1a of the main text (panel~\textbf{c}) and the second derivative of this photoemission intensity distribution taken in the direction of energy (panel~\textbf{d}). Solid and dashed lines are as in Fig.1a of the main text. White arrows indicate the flattening of the energy dispersion of the middle hole band due to orbital mixing induced by spin-orbit coupling.}
\end{figure}

The results of this comparison are shown in Fig.~\ref{fig:zcomp}a. The red and green solid lines are the parabolic dispersions obtained from the fit of the data points of the corresponding color (only the left and right branches of the parabola were used in the fit of the red and green data points, respectively). It is clear that the independent fit of the right (green solid line) branch gives results very similar to the simultaneous fit of both branches (blue solid line). The independent fit of the left branch deviates somewhat from the simultaneous fit, albeit it provides only a marginal improvement in the fit's figure of merit. To quantify the effect of this uncertainty on the extracted renormalization of the effective mass we have calculated the latter using the formula given in the main text, $m(E)=\hbar^2k(E)dk/dE$, independently for the left and right branches. The resulting energy-dependent effective mass is shown in Fig.~\ref{fig:zcomp}b for the left (red open circles) and right (green open circles) branches of the dispersion along with their average (blue open circles) plotted in Fig.1e of the main text. It can be seen that even the lowest value of the renormalization corresponding to the independent fit of the left branch is still significantly larger than the renormalization of all the other bands and the largest in the family of the high-temperature iron-pnictide superconductors. In view of the independent fits shown in Fig.~\ref{fig:zcomp} it is more likely that the true value of the renormalization is closer to the average than to the lowest value. Similar results can be obtained through the analysis of quasiparticle dispersion in the symmetrized (with respect to zero wave vector) data.

\begin{figure}[t!]
\includegraphics[width=\textwidth]{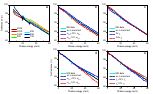}
\caption{\label{fig:refl}\textbf{Details of the far-infrared data and analysis.}~\textbf{a,} Photon-energy dependence of the far-infrared reflectance at various temperatures.~\textbf{b--e,} Analysis of the reflectance at $30\ \textrm{K}$ (cyan solid line) in the framework of the two-Drude model (dotted lines). Black dotted line shows the reflectance corresponding to the best-fit parameters of the two Drude terms, as specified in Fig.1g of the main text. Red and blue dotted lines demonstrate the effect of detuning these parameters up (blue lines) or down (red lines) from the best-fit values, as detailed in the figure legends. The high-frequency dielectric function used in this model was taken to be $\varepsilon_\infty=60$, determined by the contribution of the interband transitions at higher energies.}
\end{figure}

Figures~\ref{fig:zcomp}c,d demonstrate the flattening of the middle hole band's dispersion (white arrows) near its top due to the orbital-mixing effect induced by the spin-orbit coupling.

\section{Inherent character of the orbitally selective band shifts in the electronic structure of \nfcaopt}
\par In this section we would like to briefly substantiate the bulk-related character of the orbitally selective band shifts or, equivalently, disprove their surface-related origin. Orbitally selective band width renormalization and band shifts are generic to iron-based superconductors, have been observed in essentially all materials of this family~[S~\onlinecite{PhysRevLett.105.067002S}--\onlinecite{PhysRevLett.110.167002S}]
, and suggested to result from orbital blocking in the presence of strong Hund's coupling~\cite{YinHauleKotliar_HundsCouplingAndRenorm_2011S} and a pronounced particle-hole asymmetry~\cite{Charnukha_SFCAO_2015S}, respectively. This renormalized and band-shifted electronic structure, as well as the superconducting energy gap that develops on it, observed via ARPES has been found to agree remarkably well with many bulk-sensitive techniques, strongly suggesting that the electronic band structure of the surface layer in these compounds is reflective of that in the bulk. The only exception to this rule is the 1111 type of compounds, in which the presence of both bulk- and surface-related states in the photoemission signal has been found~\cite{Shen_LaFePO_ElStr_2008S,HaiYun3761S,PhysRevLett.101.147003S,Liu2009491S,Lu2009452S,PhysRevLett.105.027001S,PhysRevB.82.075135S,PhysRevB.82.104519S,Yang2011460S,PhysRevB.84.014504S}, making the extraction of the inherently bulk electronic structure challenging. Recently, a detailed ARPES work succeeded in disentangling the bulk and surface contributions to the photoemission signal and demonstrated that the bulk electronic structure features the same orbitally selective band width renormalization and band shifts as observed in other classes of iron-based superconductors~\cite{Charnukha_SFCAO_2015S}, proving its generic and inherent character. Finally, it is well-known that 111-type materials cleave between the two layers of the intercalant, which results in a non-polar surface. The equivalence of the bulk and surface electronic states has been further demonstrated in an explicit density-functional calculation~\cite{PhysRevB.82.184518S}.

\section{Details of the far-infrared data and analysis}
\par Here we would like to demonstrate the degree of uncertainty in the extracted parameters of the free-charge-carrier response shown in Fig.1g that the finite signal-to-noise ratio of and the uncovered lowest-frequency range in the experimental data allow. We do so by comparing the optical response corresponding to the best fit shown in Figs.1f,g with that derived from the Drude-term parameters slightly detuned from those optimal values. In order to eliminate the integral transformation via Kramers-Kronig relations and to emphasize the functional shape of the reflectance assumed in the extrapolation region at lowest frequencies for Kramers-Kronig analysis, we demonstrate these trends in the raw far-infrared reflectance data. The latter are shown for different temperatures between $10\ \textrm{K}$ and $300\ \textrm{K}$ in Fig.~\ref{fig:refl}a. Figures~\ref{fig:refl}b--e show the reflectance obtained in our experiment as a function of photon energy at $30\ \textrm{K}$ overlaid with that produced by assuming two Drude terms ($\omega_{\mathrm{pl,1}}, \gamma_1$; $\omega_{\mathrm{pl,2}}, \gamma_2$) with the parameters specified in the figure legends. The high-frequency dielectric function is taken to be $\varepsilon_\infty=60$, determined by the contribution of the interband transitions at higher energies. The default (best-fit) parameters of the free-charge-carrier response referred to in these panels are those specified in Fig.1g of the main text.

\begin{SCfigure}[][b!]
\includegraphics[width=\textwidth/3]{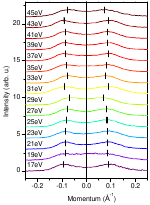}
\caption{\label{fig:dispkz}\textbf{Out-of-plane dispersion of the hole band along the $\Gamma$-Z direction in the Brillouin zone.}~Momentum-distribution curves (MDCs) at near-normal photoemission ($0$ momentum corresponds to normal emission) averaged in a $5\ \textrm{meV}$ energy window around the Fermi level for various energies of incident photons. The data were recorded in the normal state of $\textrm{NaFe}_{0.95}\textrm{Co}_{0.05}\textrm{As}$ at $27\ \textrm{K}$. The MDCs are displaced vertically by a constant of $1.5$. Black vertical lines indicate the position of the Fermi wave vector extracted as the maximum of the quasiparticle peak for every MDC.}
\end{SCfigure}

\section{Determination of band-specific plasma frequencies based on ARPES data}
\par In optimally doped \nfcaopt\ the Fermi surface consists of three sheets: two electonlike in the corners of the Brillouin zone and one holelike at the center~\cite{PhysRevB.86.214508S}. It is well-known and a generic feature of iron-based superconductors that the bands generating the electron sheets are only weakly dispersive in the out-of-plane direction, giving rise to two quasi-cylindrical sheets of the Fermi surface with relatively small corrugations~\cite{2009PhyC469614MS}. Therefore, we can approximate these two sheets by two identical ideal cylinders with their radii equal to the average of the long and the short half-axis of the ellipsoids giving rise to the corrugation. The electronic band generating the hole sheet of the Fermi surface also disperses rather weakly along the $\Gamma$-Z direction of the Brillouin zone, as shown in Fig.~\ref{fig:dispkz}: the Fermi wave vector of this band remains within $0.07$--$0.09$~\AA$^{-1}$\ between $\Gamma$ and~Z. Therefore, the Fermi-surface sheet generated by this band can likewise be approximated by a cylinder with a radius of about $0.08$~\AA$^{-1}$. Thus the task of calculating the plasma frequency has been reduced to a two-dimensional problem. In general, the plasma frequency of free charge carriers can be obtained from the semi-classical expression for the dielectric function~\cite{Cardona_semiconductors_articlestyleS} assuming the intraband case and low temperatures and is given by the following expression (in SI units):
\begin{equation}
\omega_{\mathrm{pl},\sigma}^2=\frac{e^2}{8\pi^3\hbar^2\varepsilon_0}\sum_n\int\left[\frac{dE_{n\mathbf{k}}}{dk_\sigma}\right]^2\delta(E_{n\mathbf{k}}-E_{\mathrm{F}})d^3\mathbf{k},\label{eq:plgen}
\end{equation}
where $\sigma=x,y,z$ enumerates the coordinate axes, $n$ is the band index, $\delta(x)$ is Dirac's delta function, $E_{n\mathbf{k}}$ is the energy dispersion of the $n$th band, $E_{\mathrm{F}}$ is the Fermi energy, $e$ is the electron charge, $\hbar$ is the reduced Planck constant, and $\varepsilon_0$ is the vacuum permittivity. Assuming our two-dimensional approach and the tetragonal symmetry of \nfcaopt, $\omega_{\mathrm{pl},z}=0$ and $\omega_{\mathrm{pl},x}=\omega_{\mathrm{pl},y}=\omega_{\mathrm{pl}}$. Thus only one quantity $\omega_{\mathrm{pl}}$ needs to be calculated for each band. In the two-dimensional case and for an isotropic parabolic band with an effective band mass $m_{\mathrm{eff}}$ [$E_{\mathbf{k}}=\hbar^2\mathbf{k}^2/(2m_{\mathrm{eff}})$] Eq.~\ref{eq:plgen} can be rewritten as a simple and familiar expression:
\begin{equation}
\omega_{\mathrm{pl}}^2=\frac{N_{\mathrm{2D}}(E_{\mathrm{F}})e^2}{m_{\mathrm{eff}}\varepsilon_0},\label{eq:pl2d}
\end{equation}
where $N_{\mathrm{2D}}(E_{\mathrm{F}})=\frac{k_{\mathrm{F}}^2}{(2\pi)^2}\frac{2\pi}{c}$ is the density of states at the Fermi level for an isotropic parabolic band in two dimensions, $k_{\mathrm{F}}$ is the Fermi wave vector, and $c$ is the out-of-plane lattice constant. Thus the calculation of the band-specific plasma frequencies is reduced to the determination of the Fermi wave vector and the quasiparticle effective mass for each band, which can easily be done based on the photoemission intensity distributions near the Fermi level presented in Figs.1a,b. As already mentioned above and shown in Fig.~\ref{fig:dispkz}, the Fermi wave vector of the hole band is about $0.08$~\AA$^{-1}$, and its effective mass is found to be $3.83\ m_{\mathrm{e}}$ (where $m_{\mathrm{e}}$ is the bare electron mass), which results in a plasma frequency of $\omega_{\mathrm{pl}}^{\mathrm{hole,1}}=0.23\ \textrm{eV}$. The Fermi wave vectors and effective masses of the electron bands at the M-point are $0.086$~\AA$^{-1}$, $0.92\ m_{\mathrm{e}}$ and $0.153$~\AA$^{-1}$, $2.34\ m_{\mathrm{e}}$ for the inner and the outer band, respectively. These values result in almost identical plasma frequencies of $\omega_{\mathrm{pl}}^{\mathrm{el,in}}=0.5\ \textrm{eV}$ and $\omega_{\mathrm{pl}}^{\mathrm{el,out}}=0.56\ \textrm{eV}$, respectively.
\newpage

\end{document}